\documentclass[12pt]{article} 
\usepackage{graphicx}

\newcommand\pubnumber{DPF2013-92}
\newcommand\pubdate{\today}

\def\napoli{Department of Physics\\
University of Arizona, Tucson, AZ, USA}

\def\Title#1{\begin{center} {\Large #1 } \end{center}}
\def\Author#1{\begin{center}{ \sc #1} \end{center}}
\def\Address#1{\begin{center}{ \it #1} \end{center}}

\newcommand\pubblock{\rightline{\begin{tabular}{l} \pubnumber\\
         \pubdate  \end{tabular}}}
\newenvironment{Abstract}{\begin{quotation}  }{\end{quotation}}
\newenvironment{Presented}{\begin{quotation} \begin{center} 
             PRESENTED AT\end{center}\bigskip 
      \begin{center}\begin{large}}{\end{large}\end{center} \end{quotation}}

\def\beq{\begin{equation}}
\def\eeq#1{\label{#1}\end{equation}}
\def\eeqn{\end{equation}}

\def\beqa{\begin{eqnarray}}
\def\eeqa#1{\label{#1}\end{eqnarray}}
\def\eeqan{\end{eqnarray}}

\def\sst{\scriptscriptstyle}

\let\bar=\overbar

\def\Dslash{\not{\hbox{\kern-4pt $D$}}}
\def\dslash{\not{\hbox{\kern-2pt $\del$}}}

\def\msb{{\bar{\ssstyle M \kern -1pt S}}}

\usepackage{atlasphysics} 
\usepackage{url}
\usepackage{subfigure}
\usepackage{hyperref}

\begin{document}
\begin{titlepage}
\pubblock

\vfill
\Title{Search for anomalous production of events with same-sign dileptons
  and $b$ jets  in 14.3 fb$^{-1}$ of pp collisions at $\sqrt{s} = 8$ TeV
  with the ATLAS detector}
\vfill
\Author{Xiaowen Lei} 
\Address{\napoli}
\vfill
\begin{Abstract}
  A search is presented for exotic processes that result in final states
  containing jets including at least one $b$ jet, sizable missing transverse
  momentum, and a pair of leptons with the same electric charge.
  Using a sample of 14.3~\ifb{} of $pp$ collisions at $\sqrt{s} =8$~\TeV{}
  recorded by the ATLAS detector at the Large Hadron Collider, no
  significant excess of events over the background expectation is observed.
  This observation is interpreted as constraining several signal hypotheses
  beyond the Standard Model and limits are set at 95\% confidence level on
  relevant parameters of the signal hypotheses.
\end{Abstract}
\vfill
\begin{Presented}
DPF 2013\\
The Meeting of the American Physical Society\\
Division of Particles and Fields\\
Santa Cruz, California, August 13--17, 2013\\
\end{Presented}
\vfill
\end{titlepage}
\def\thefootnote{\fnsymbol{footnote}}
\setcounter{footnote}{0}

\section{Introduction}

The Standard Model (SM) has been repeatedly confirmed experimentally.
Nonetheless there appears to be a need for physics beyond the SM at about the
weak scale, with additional features that explain the presence of dark matter
in the universe, and provide a mechanism to naturally stabilize the Higgs boson
mass. This note reports on a search for new physics resulting in isolated
high-$p_T$ lepton pairs with the same electric charge and $b$ jets. This is a
promising search channel since the SM yields of such events are small, and
several types of new physics may contribute. The analysis is described in
detail in~\cite{TheATLAScollaboration:2013jha}.

Among the models that predict enhanced same-sign lepton production are those
that postulate the existence of a fourth generation of chiral quarks, the
existence of vector-like quarks (VLQ), production of two positively-charged
top quarks, and an enhancement of the four top quark production cross
section. We use a common data sample to search for each of these signatures,
and optimise the event selection criteria for each signal model.

This analysis studies the chiral fourth generation quark with charge $-1/3$,
called the $b^\prime$~\cite{Aad:2012bb}. There are several possible
varieties of VLQ. This analysis studies those with the same electric charge
as the SM $b$ and $t$ quarks, called the $B$ and $T$. It is assumed for this
analysis that VLQ decay predominantly to third-generation SM quarks. The
branching fractions to each allowed final state are model-dependent. As a
reference we will use the branching ratios naturally occurring in models
where the $B$ and $T$ exist as singlets or
doublets~\cite{AguilarSaavedra:2009es}.

The same-sign top pair production can be described by an effective four-fermion
contact interaction, with separate operators for the different initial-state
chiralities~\cite{Aad:2012bb}. The four top quarks production can generically
be described in terms of a four-fermion contact interaction with coupling
strength $C/\Lambda^2$~\cite{Degrande:2010kt}. Two specific models are also
considered. The first is sgluon pair production, where sgluons are
color-adjoint scalars. The second model is one with two universal extra
dimensions under the real projective plane geometry (2UED/RPP)~\cite{Lyon09}.
This model predicts the pair production of Kaluza-Klein excitations of the
photon with mass $\sqrt{2}m_{KK}$.

\section {Data and Monte Carlo Simulation}

The data used were recorded by the ATLAS detector~\cite{Aad:2008zzm} at the LHC
$pp$ collider operating at $\sqrt{s} = 8$ \TeV{} between April and October
2012, and correspond to an integrated luminosity of 14.3~\ifb{}. Signal and
some background sources were modelled using Monte Carlo (MC) simulations. The
remaining background sources are determined from the data, as described in
Section~\ref{sec:bkg}.
 
\section{Event selection}

The final state considered in this search requires the presence of exactly
two isolated leptons \footnote{Only electrons and muons are considered in
  the search. Tau leptons are not explicitly reconstructed, but electrons
  and muons from tau decay may enter the selected samples.} in the event,
with the same electric charge. In addition, two or more jets are required,
at least one of which is consistent with being a $b$ jet, and sizable
missing transverse momentum (\met{}). Other event selection criteria include
at least one good primary vertex and passing either an electron or muon
trigger. Data events are also required to be from a run where the detector
performance is known to be good.

Depending on the flavour of the leptons the event is identified as an $ee$,
$e\mu$, or $\mu\mu$ candidate.  If the same-sign leptons are of the same
flavour, their invariant mass $m_{\ell\ell}$ is required to be $> 15$ \GeV{}
and to satisfy $|m_{\ell\ell}-m_Z| > 10$ \GeV{}. This rejects events from a
resonance where the charge of one lepton is misidentified. Finally, the scalar
sum of all jet and lepton \pt's (\HT) is required to be $>550$ \GeV{}. These
basic selection criteria are applied to all searches; some of them are
tightened when optimising the selection for each signal model (see
Section~\ref{sec:opt}).

\section {Background Estimation}
\label{sec:bkg}

Backgrounds arise from two distinct sources: SM processes that result in the
same final-state objects as the signal, and instrumental backgrounds where
objects are misidentified or misreconstructed such that they appear to have
the same final state. The former category is estimated using MC simulation.
The latter category can be further separated into two categories: $i$)
events that contain two leptons of opposite charge, where one of the charges
is mismeasured and $ii$) events where one or more jets are misidentified as
leptons.

To estimate the number of events with misidentified leptons (fakes), we
begin by relaxing the lepton identification criteria to form a
\textit{loose} sample.  Single lepton events were used to measure the
efficiencies $r$ and $f$ for real and fake leptons selected using the loose
criteria to also satisfy the standard (or \textit{tight}) criteria. The
estimated number of events with misidentified leptons in the selected sample
is then estimated using a matrix method with a four-by-four matrix. Care
must be taken in applying the procedure to muons, since the difference
between the tight and loose criteria for muons is only in the isolation
requirement, and the trigger used for low-\pt\ muons also requires
isolation. This is addressed by requiring low-\pt\ muons in the loose sample
to satisfy isolation criteria similar to those applied in the trigger, and
to apply different values of $r$ and $f$ for muons that require this
isolation.

Charge misidentification is negligible for muons due to the presence of the
muon system. For electrons, the rate of charge mismeasurement
($\varepsilon$) is first calculated from a sample of $Z \ra ee$ events. The
rate is parametrized in electron $\eta$ and measured for several regions of
electron \pt. In the high \pt\ region, $\varepsilon$ is scaled to the value
measured using $t\bar{t}$ MC simulated events, in order to reduce the large
uncertainty due to limited $Z \ra ee$ statistics. To determine the number of
events expected from charge mismeasurement in the signal region, a sample is
selected using the same criteria as for the analysis selection, except that
an opposite-sign rather than same-sign $ee$ or $e\mu$ pair is required.  The
measured $\varepsilon$ values are then applied to each electron in this
sample to determine the charge misidentification yield in the analysis
sample.

To correct for the overlap between the charge misidentification and fake
electron estimates, the charge mismeasurement rate is recalculated with
the estimated fake electron contribution removed from the tight selection
sample. The ratio of the initial and recalculated rates is taken as an
$\eta$- and \pt-dependent correction factor to be applied to the charge
mismeasurement rates.

The background estimates are validated using samples selected with criteria
similar to our standard selection, but where the expected yield from signal
events is small. One such control region (called the \met{} control region)
is defined by dropping the requirement on \met{} and requiring
$100~<~\HT~<~400$~\GeV{}. The predicted and observed yields in the \met{}
control region are given in Table~\ref{ctrl:met_lowht_yield}.

\begin{table}[tt]
  \begin{center}
    \caption{Observed and expected number of events with 
        statistical (first) and systematic (second) uncertainties for the \met{} control region
        selection. For the Monte Carlo simulation, the systematic uncertainties
    include only the production cross section uncertainty.}\label{ctrl:met_lowht_yield}
    \begin{tabular}{l|c|c|c}
      \hline\hline
       & \multicolumn{3}{c}{Channel} \\
      \cline{2-4}
      Samples & $ee$ & $e\mu$ & $\mu\mu$ \\
      \hline
      Charge misid & $25.7\pm 0.7 \pm 6.6$ & $30.2\pm 0.6 \pm 7.9$ & --- \\
      Fakes & $38.7\pm 3.7 \pm 11.6$ & $73.1 \pm 5.3 \pm 21.9$ & $33.4 \pm 8.5 \pm 10.0$ \\
      \hline
      Diboson & & & \\
      $\bullet$ $WZ/ZZ$+jets & $3.9\pm 0.7 \pm 1.3$ & $10.9\pm 1.2\pm 3.7$ & $5.1\pm 0.8\pm 1.7$ \\
      $\bullet$ $W^{\pm}W^\pm$+2 jets & $0.4\pm 0.2\pm 0.2$ & $1.2\pm 0.3\pm 0.6$ & $0.8\pm 0.2\pm 0.4$ \\
      \hline
      $t\bar{t}+W/Z$ & & & \\
      $\bullet$ $t\bar{t}W$(+jet) & $1.7\pm 0.1\pm 0.5$ & $6.6\pm 0.2\pm 2.0$ & $4.3 \pm 0.2\pm 1.3$ \\
      $\bullet$ $t\bar{t}Z$(+jet) & $0.5\pm 0.1\pm 0.1$ & $1.5\pm 0.1\pm 0.5$ & $0.8\pm 0.1\pm 0.2$ \\
      $\bullet$ $t\bar{t}W^+W^-$ & $0.014\pm 0.002$ & $0.050\pm 0.004$ & $0.029\pm 0.003$ \\
      \hline
      Total expected bkg & $71 \pm 5 \pm 13 $ & $124 \pm 8 \pm 24$ & $44 \pm 11 \pm 10$ \\
      \hline
      Observed & 64 & 97 & 38 \\
      \hline
      Signal contamination & & & \\
     $\bullet$ b$^\prime\rightarrow Wt$ (800 \GeV{}) & $< 0.003$   & $0.009 \pm 0.006$   &
$0.002 \pm 0.001$ \\
    $\bullet$ 4 tops contact & $0.009\pm0.005$ & $0.06\pm0.02$ & $0.02\pm0.01$ \\
    ($C/\Lambda^2 = -4\pi \TeV{}^{-2}$) & & & \\
      \hline
    \end{tabular}
  \end{center}
\end{table}

\section{Systematic Uncertainties} 
\label{sec:syst}

The expected signal and background yields are subject to several systematic
uncertainties. For the yields derived from simulation, the largest source of
uncertainty is that on the calculated cross section. The largest
uncertainties specific to ATLAS arise from the jet energy scale,
the $b$-tagging efficiency, and the lepton identification efficiency.

Systematic uncertainties on the backgrounds estimated from data are
evaluated separately. The uncertainties on the charge mis-identification
rates and correction factors are propagated to the systematic uncertainty on
the charge mis-identification prediction. The level of agreement between the
observed and predicted yields in the control regions is taken as an estimate
of the systematic uncertainty in the fake lepton prediction, resulting in a
30\% uncertainty.

The sources of systematic uncertainties that contribute more than 1\%
uncertainty on the expected signal or background yield for the
$b^\prime$/VLQ selection are summarised in Table~\ref{tab:syst}. These
uncertainties have similar impact on the expected yields for the other
signal models.
 
\begin{table}[tt]
  \begin{center}
    \caption{Leading sources of systematic uncertainty on the signal and
      background estimates for the $b^\prime$/VLQ selection, and their
      relative impact on the total background estimate.  A $b^\prime$ mass
      of 650 \GeV{} is assumed.}\label{tab:syst}
    \begin{tabular}{l|ccc|ccc}
      \hline\hline
       & \multicolumn{6}{c}{Uncertainty in \%} \\
       \hline
       &\multicolumn{3}{c}{650 \GeV{} $b^\prime$} & \multicolumn{3}{c}{Background} \\
       \hline
      Source & $ee$ & $e\mu$ & $\mu\mu$ & $ee$ & $e\mu$ & $\mu\mu$ \\
      \hline
      Cross section  & -- & -- & -- & 14.4  & 25.4 & 32.9 \\
      Fakes      & -- & -- & -- & 9.7 & 1.4 & 10.1 \\
      Charge misidentification & -- & -- & -- & 7.2 & 7.1 & -- \\
      Jet energy scale & 4.6 &  2.5 & 0.2 & 3.5 & 10.2 & 4.4 \\
      ISR/FSR & 6.0 & 6.0 & 6.0 & 2.6 & 4.5 & 4.0 \\
      $b$-tagging efficiency &4.6 & 3.1 & 3.0 & 2.1 & 4.4 & 4.0 \\
      Lepton ID efficiency & 5.3 & 4.9 & 8.2 & 2.2 & 3.6 & 5.4 \\
            Jet energy resolution & 0.8 & 0.9 & 0.3 & 0.9 & 2.7 & 2.0 \\
      Luminosity & 3.6 & 3.6 & 3.6 & 1.6 & 2.7 & 3.6 \\
      Lepton energy scale & 0.8 & 0.4 & 0.0  & 1.4 & 0.9 & 0.1 \\
      JVF selection efficiency & 2.5 & 2.9 & 2.6 & 1.1 & 1.5 & 1.4 \\
       \hline
    \end{tabular}
  \end{center}
\end{table}

\section{Selection optimisation}
\label{sec:opt}

For each signal model, the event selection was optimised using a grid search
of requirements on \HT, \MET, the number of jets, and the number of
$b$-tagged jets. The set of criteria that provided the best expected
sensitivity to new physics (including the effect of systematic
uncertainties) was chosen. A summary of the optimised selection for each
channel is given in Table~\ref{tab:cuts}.   

\begin{table}[p]
  \begin{center}
    \caption{Summary of the selection for the various signals. For all signals, the number of jets
    is required to be at least two and $\met$ is required to be $>40$~\GeV{}.}\label{tab:cuts}
    \begin{tabular}{l|c|c|c}
      \hline\hline
      Variable & $b^\prime$ and VLQ & $tt$ & $t\bar{t}t\bar{t}$ \\
      \hline
      \HT{} & $>650\GeV{}$ & $>550\GeV{}$ & $>650\GeV{}$ \\
      $N_{b-jets}$ & $\geq 1$ & $\geq 1$ & $\geq 2$ \\
      Charge & $\pm\pm$ & $++$ & $\pm\pm$ \\
      \hline
    \end{tabular}
  \end{center}
\end{table}

\section{Results}

For each signal selection, the yield in the $ee$ and $\mu\mu$ channels is
consistent with the background expectation. There is an excess of events in
the $e\mu$ channel, which is most significant for the four top quark
selection where six events are observed. The probability of observing six or
more events, given the expected background and its statistical and
systematic uncertainties, is 3.9\%.  Since this probability is not small
enough to support any claim of new physics, we interpret the data as
constraining each of the new physics models, and report the 95\% C.L. limits
relevant for each model. The $CL_s$ method~\cite{Junk:1999kv} is used to set
the limits.

The 95\% C.L. upper limits on the $b^\prime$ pair production cross section are
shown in Fig.~\ref{limit:bprime}, as a fuction of $b^{\prime}$ mass. Comparison
with the calculated cross section implies that the observed lower limit on the
$b^\prime$ mass must be $>0.72$~\TeV{}, assuming 100\% branching fraction to
$Wt$. The corresponding expected lower limit is $0.77$~\TeV{}, and the observed
result is consistent with this expectation at about the one standard deviation
level. If light quark decays are also considered, the limits become weaker. A
two-dimensional scan of excluded $b^\prime$ masses as a function of branching
fraction to $Wt$ is shown in Fig.~\ref{limit:bprime2D}.

\begin{figure}[p]  
  \begin{center}
    \subfigure[\label{limit:bprime}]{\includegraphics[width=0.49\textwidth]{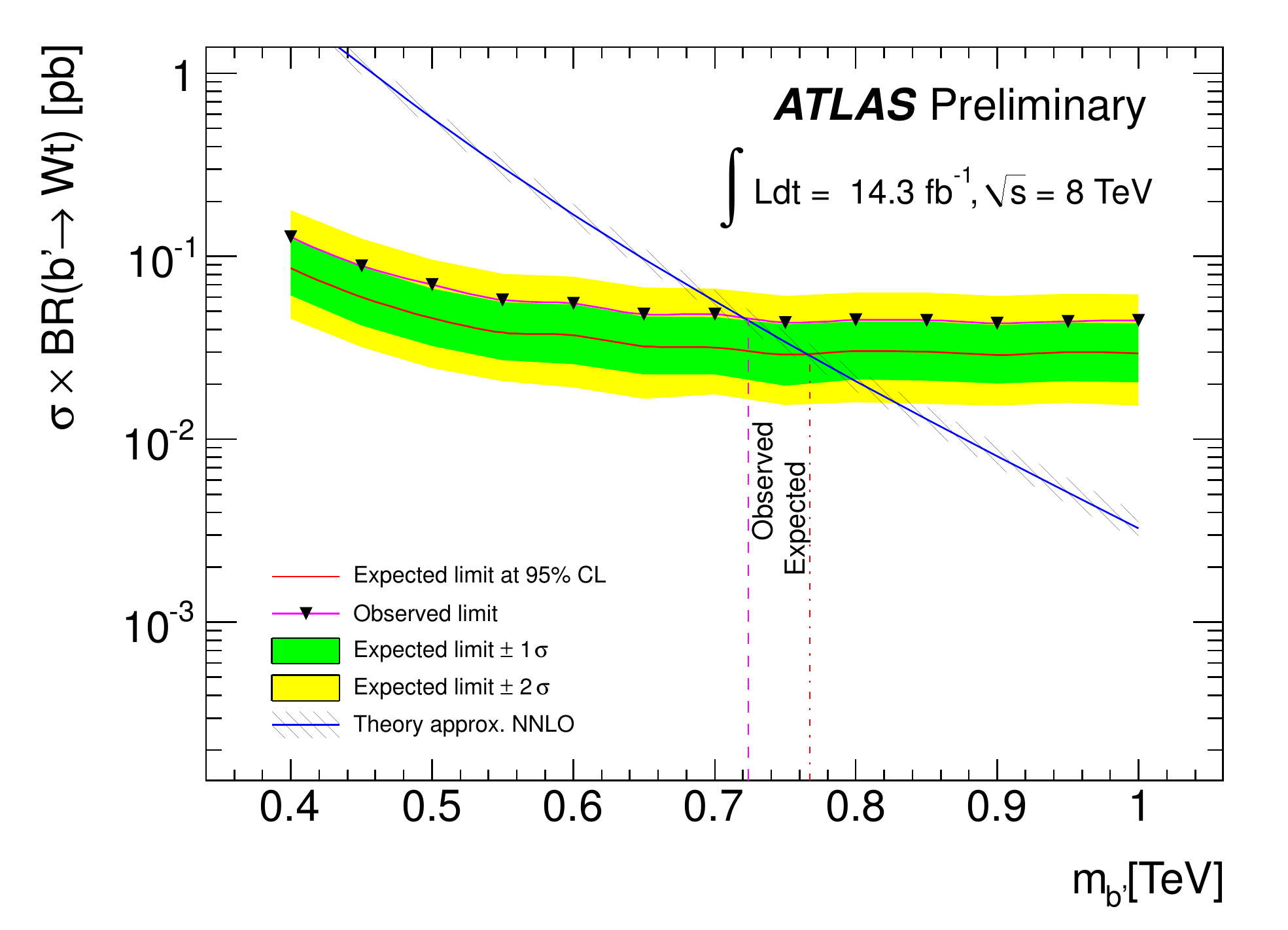}}
    \subfigure[\label{limit:bprime2D}]{\includegraphics[width=0.49\textwidth]{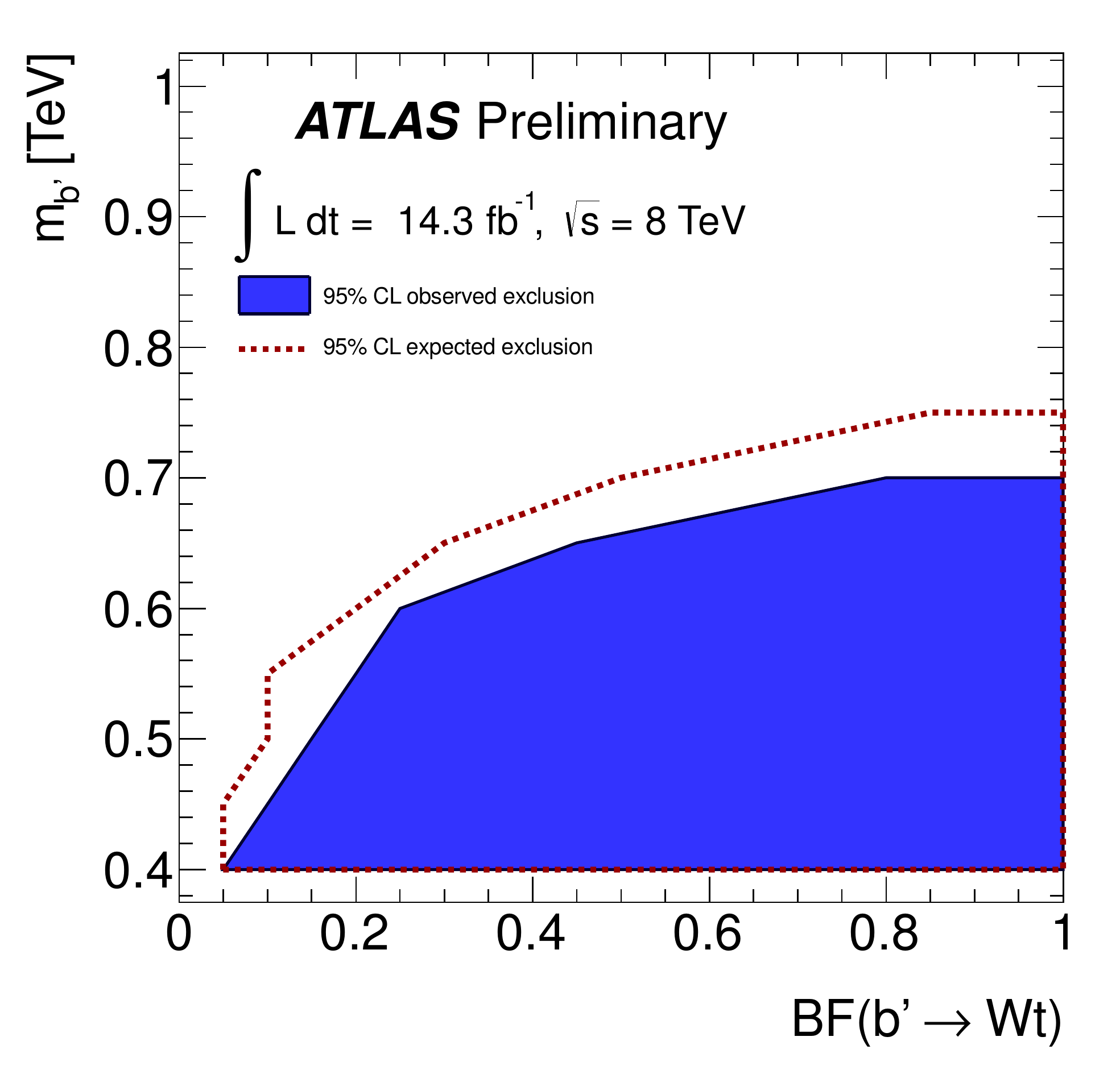}}
    \caption{Expected and observed upper limits on the pair production cross
      section of  $b^\prime\ra Wt$, (a) as a function of the $b^\prime$
      mass, and (b) as a function of the $b^\prime$ mass and the branching ratio to $Wt$,
        calculated from the $b^\prime\ra Wq$ sample}
  \end{center}
\end{figure}

Limits on the VLQ pair-production cross section, assuming the branching
ratios to $W$, $Z$, and $H$ modes prescribed by the singlet model, are shown
in Fig.~\ref{fig:limitObs:VLQ}. Comparison with the calculated cross section
results in lower limits on the $B$ mass of $0.59$~\TeV{} and on the $T$ mass
of $0.54$~\TeV{} at 95\% C.L. The expected limits are 0.63~\TeV{} for the
$B$ mass and 0.59~\TeV{} for the $T$ mass. If the three branching fractions
are allowed to vary independently, the data can be interpreted as excluding
at 95\% C.L. some of the possible sets of branching ratios for a given $B$
or $T$ mass. These exclusions are shown in Figs.~\ref{fig:BBS_2DLimits} and
\ref{fig:TTS_2DLimits}.

\begin{figure}[p]  
  \begin{center}
    \subfigure[\label{limit:sglObs}]{\includegraphics[width=0.49\textwidth]{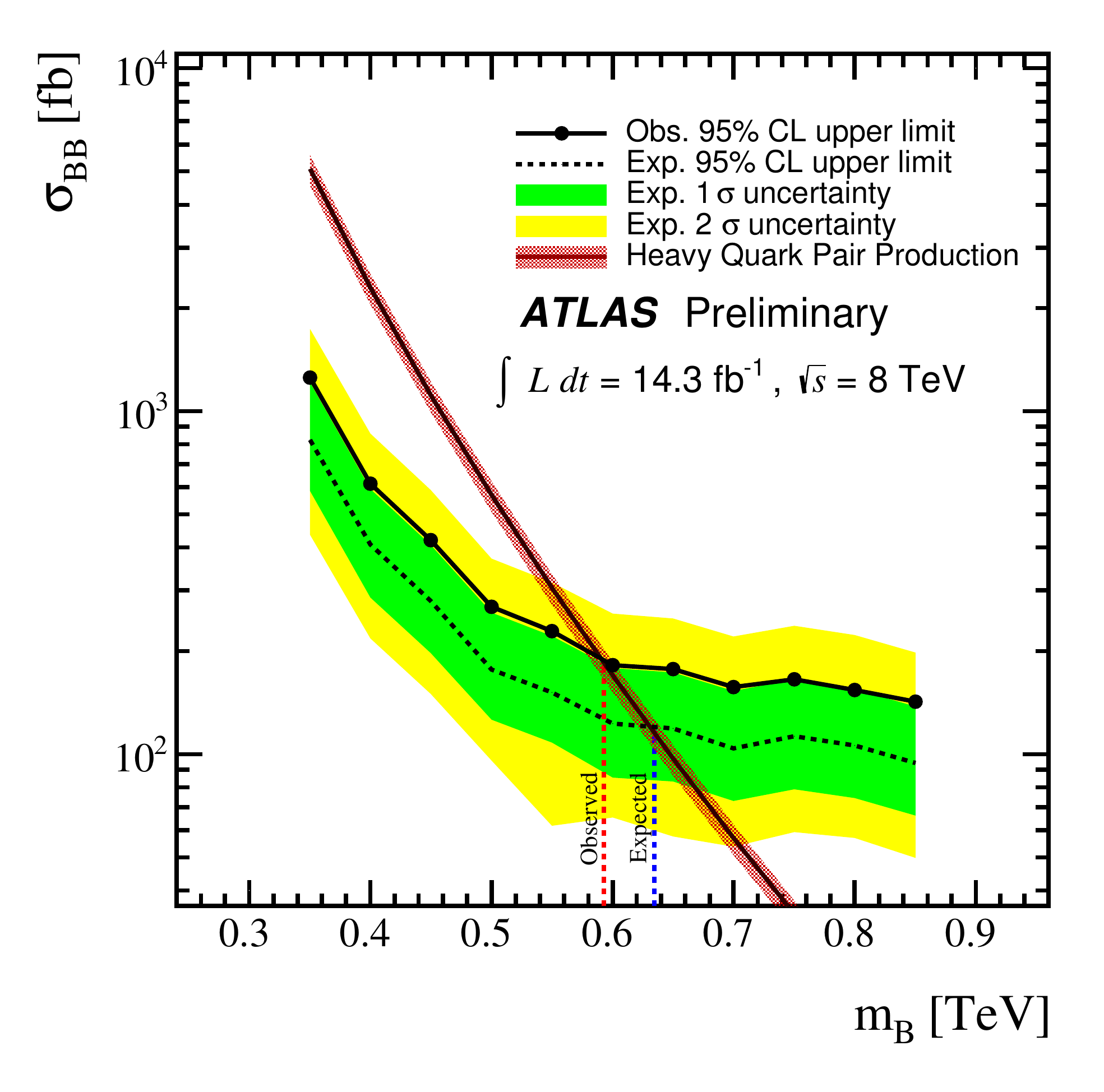}}
    \subfigure[\label{limit:RPPobs}]{\includegraphics[width=0.49\textwidth]{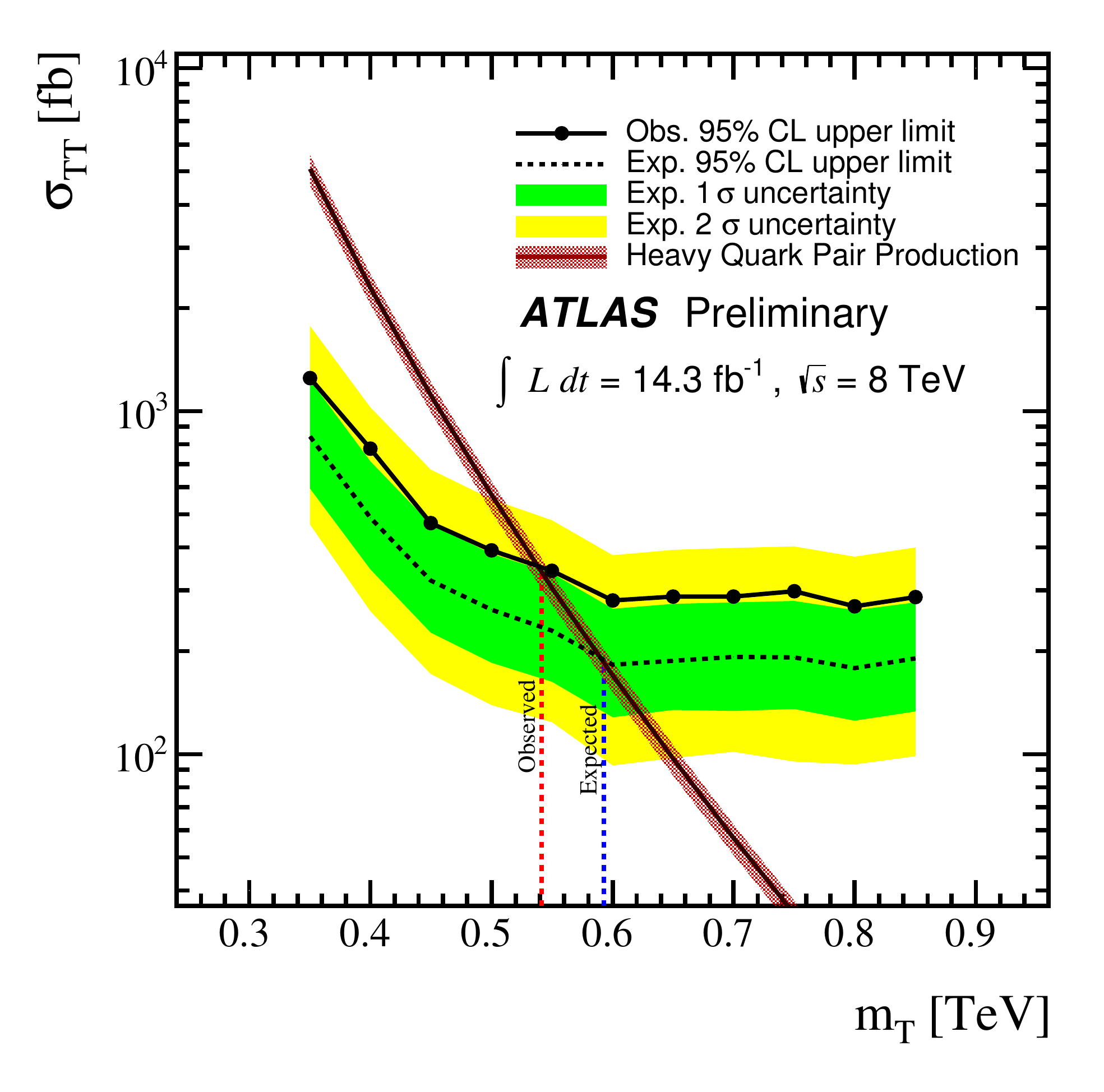}}
    \caption{Observed limits on the mass of (a) vector-like $B$ and (b)
      vector-like $T$ quarks.  These limits assume pair production, with
      branching ratios given by model where the $B$ and $T$ quarks exist as
      singlets~\cite{AguilarSaavedra:2009es}.} \label{fig:limitObs:VLQ}
  \end{center}
\end{figure}

\begin{figure}[p]  
  \begin{center}
    \includegraphics[width=0.8\textwidth]{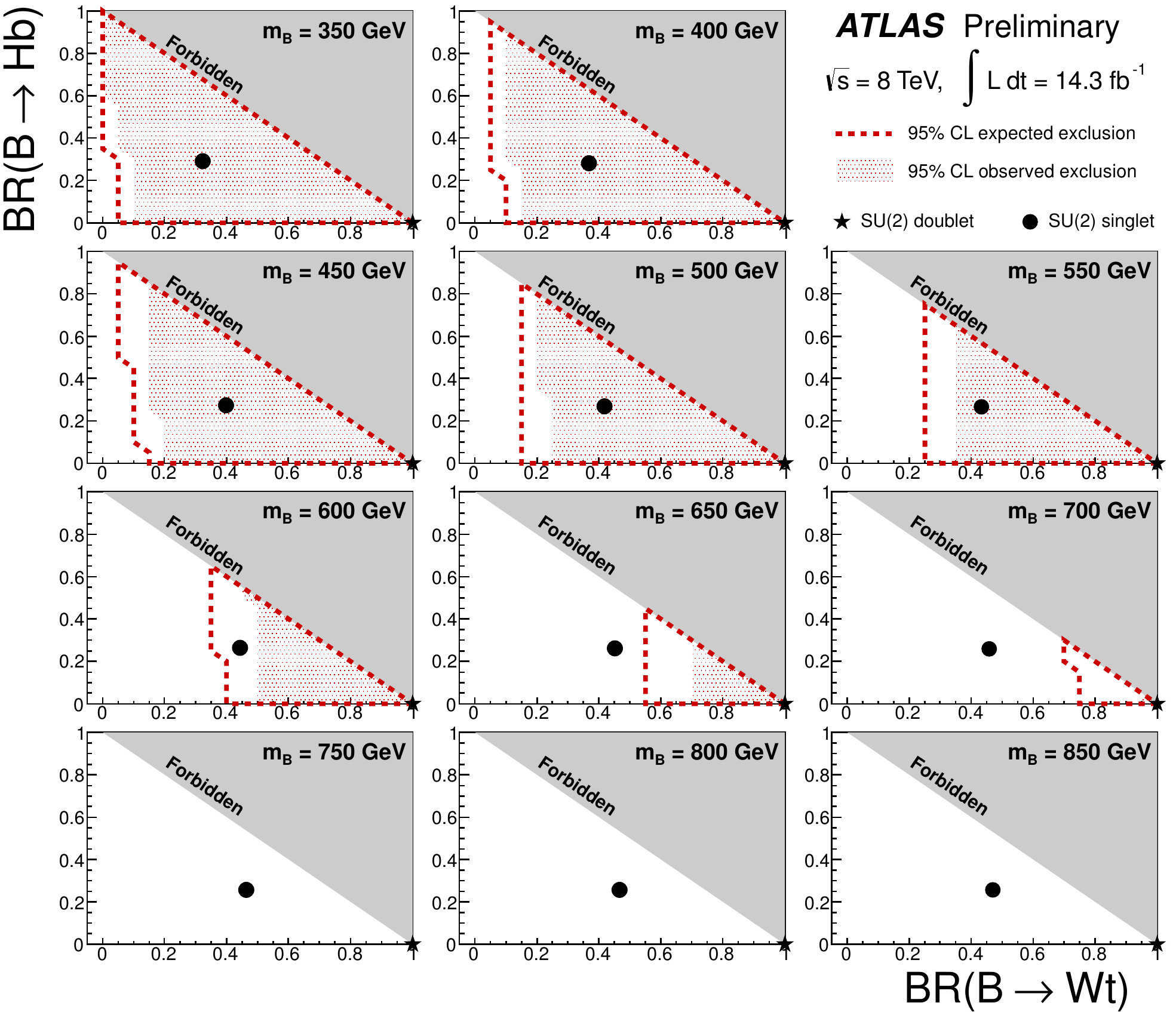}
    \caption{Observed and expected vector-like $B$ branching ratios excluded
      at 95\% C.L. for several $B$ mass hypotheses.  For reference, the
      branching ratios expected in models~\cite{AguilarSaavedra:2009es}
      where the $B$ is in a SU(2) singlet (doublet) are indicated by a
      circle (star).}
    \label{fig:BBS_2DLimits}
  \end{center}
\end{figure}

\begin{figure}[p]  
  \begin{center}
    \includegraphics[width=0.8\textwidth]{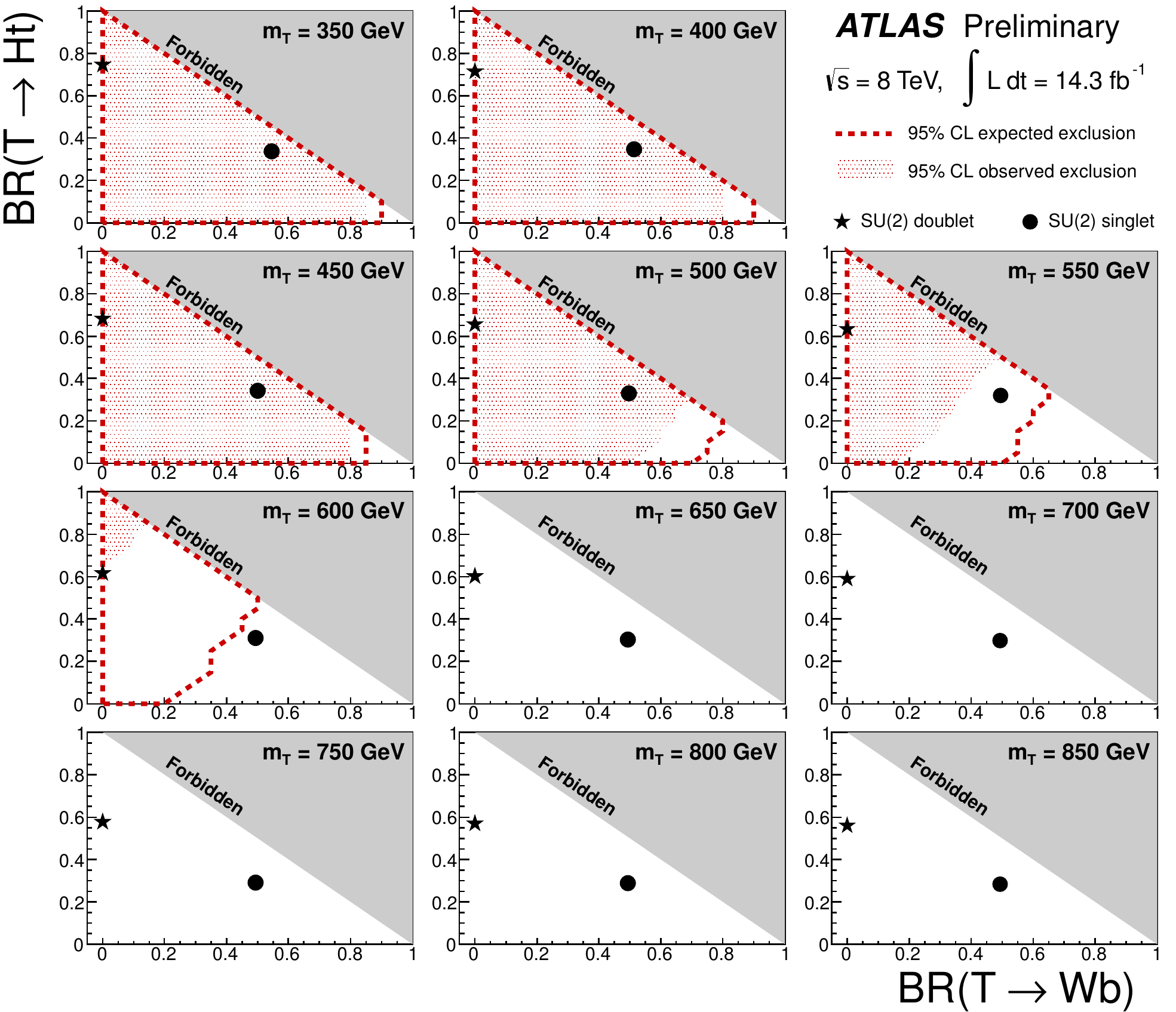}
    \caption{Observed and expected vector-like $T$ branching ratios excluded
      at 95\% C.L. for several $T$ mass hypotheses. For reference, the
      branching ratios expected in models~\cite{AguilarSaavedra:2009es}
      where the $T$ is in a SU(2) singlet (doublet) are indicated by a
      circle (star).} \label{fig:TTS_2DLimits}
  \end{center}
\end{figure}

For the positively-charged top quark pair selection, the observed upper
limit on the production cross section is 0.19~pb for left-left chirality,
0.20~pb for left-right chirality, and 0.21~pb for right-right chirality. The
corresponding expected limits are $0.19^{+0.09}_{-0.05}$ pb,
$0.21^{+0.09}_{-0.06}$ pb, and $0.22^{+0.10}_{-0.07}$ pb, respectively. For
each chirality, the upper limit on $C$ as a function of $\Lambda$ is shown
in Fig.~\ref{fig:limitObs:sstop}.

\begin{figure}[p]  
  \begin{center}
    \subfigure[]{\includegraphics[width=0.49\textwidth]{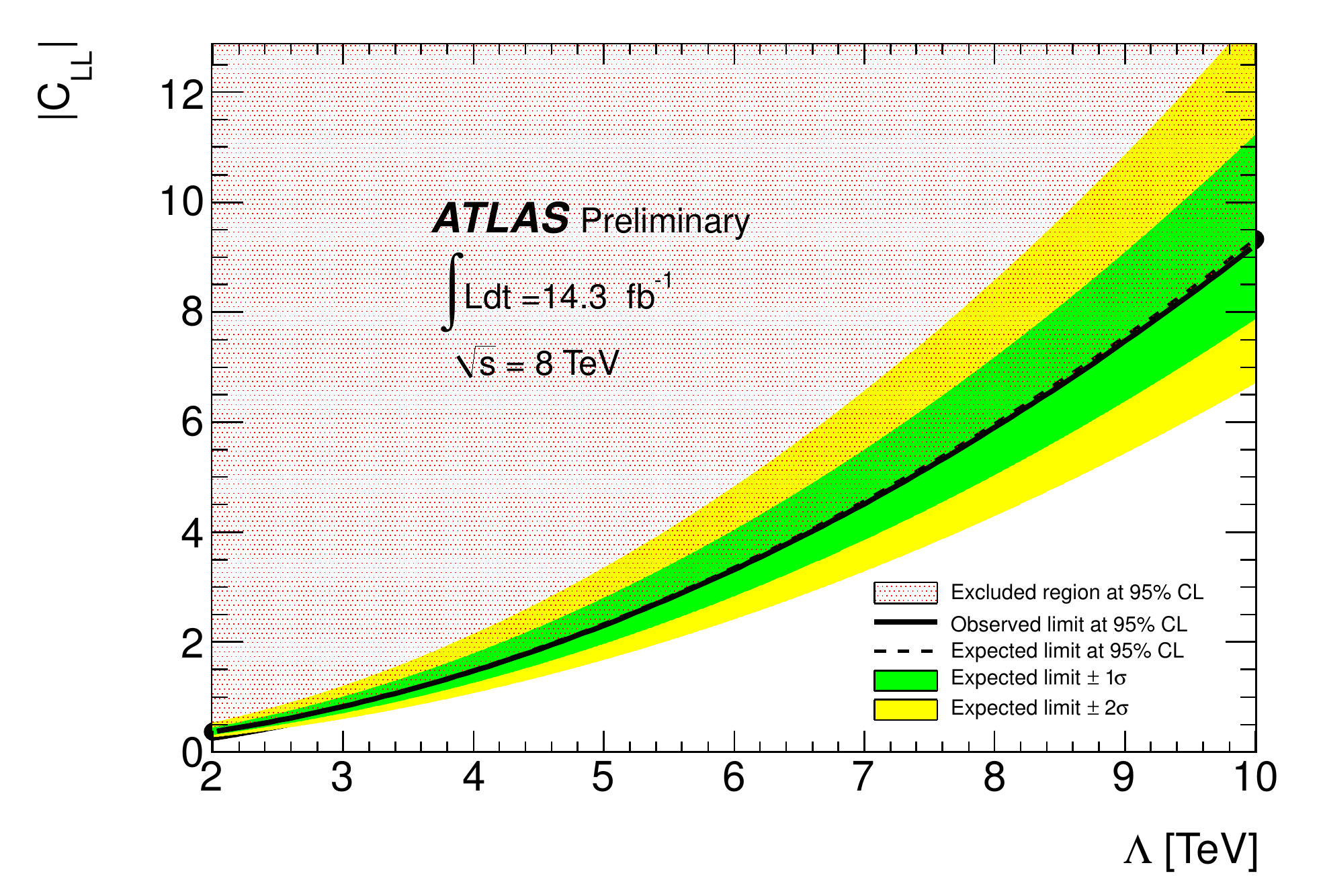}}
    \subfigure[]{\includegraphics[width=0.49\textwidth]{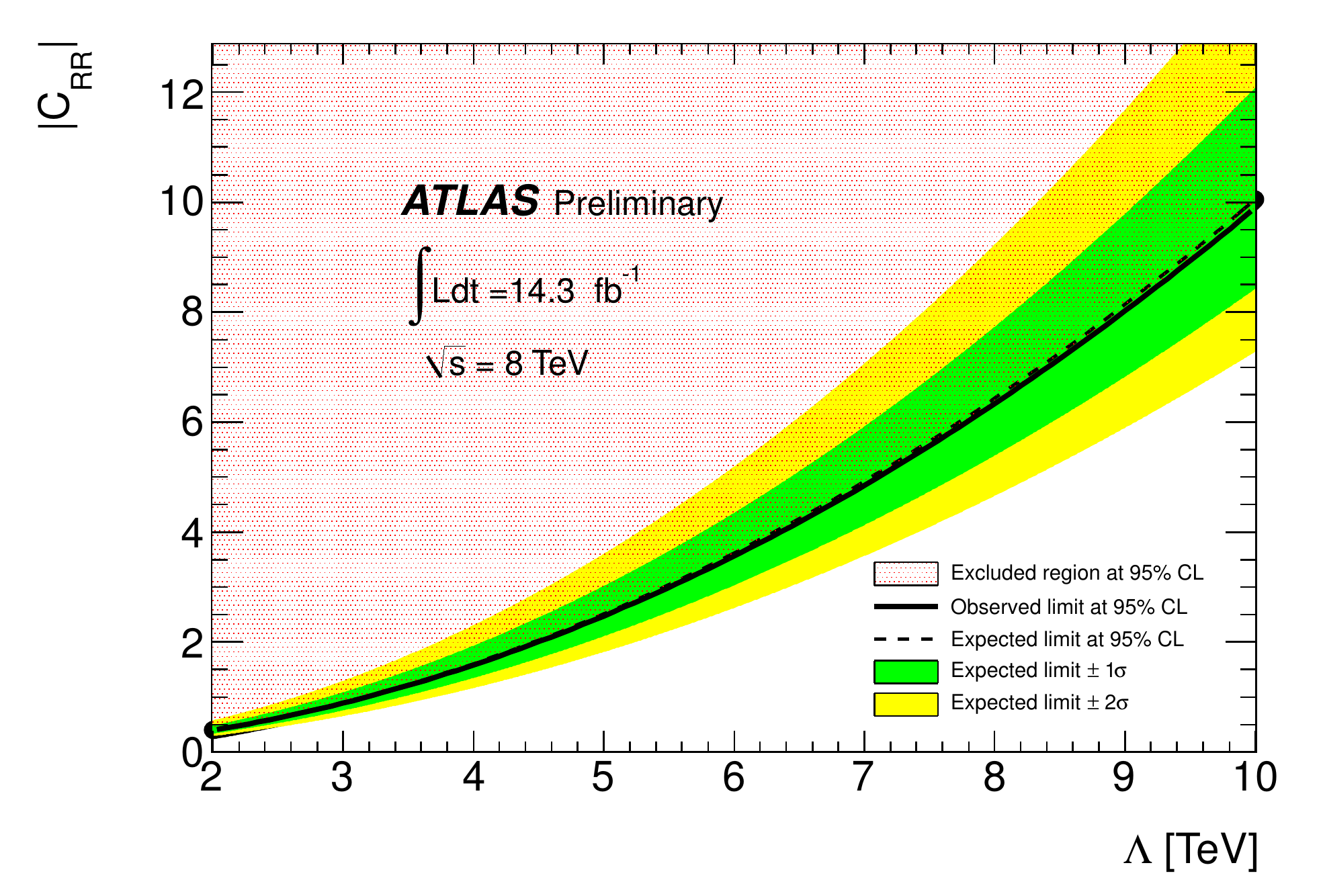}}
    \subfigure[]{\includegraphics[width=0.49\textwidth]{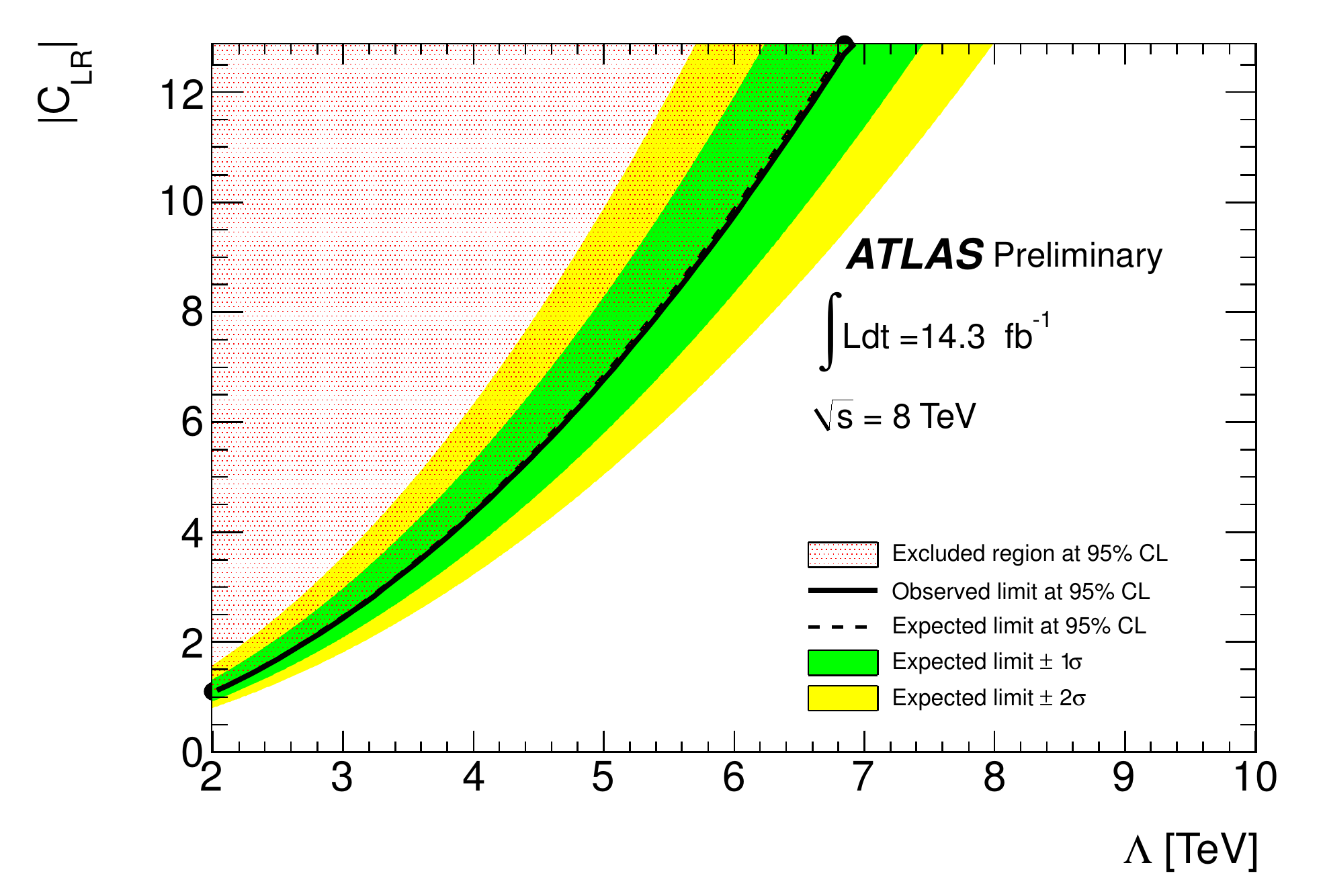}}
    \caption{Limits on positively-charged top quark pair production on the
      the coupling constant $C$ as a function of the new physics energy
      scale $\Lambda$ for (a) left-left, (b) right-right, and (c) left-right
      chiralities. The hashed region is excluded at 95\%
      C.L.}\label{fig:limitObs:sstop}
  \end{center}
\end{figure}

For the four top production, the observed upper limit on the cross section
is 85 fb assuming SM kinematics, and 59 fb for a new-physics contact
interaction. Both of these results are consistent with the expected limits
within one standard deviation. The upper limit on $C$ as a function of
$\Lambda$ is shown in Fig.~\ref{fig:limitObs:4t}.
Figure~\ref{fig:limitObs:4t} also shows the limits for the two specific new
physics models. We find that the lower limit on the sgluon mass is
0.80~\TeV{}(for an expected limit of 0.83~\TeV{}). The lower-limit on the
Kaluza-Klein mass is 0.90~\TeV{} (for an expected limit of 0.92~\TeV{}).

\begin{figure}[p]  
  \begin{center}
    \subfigure[]{\includegraphics[width=0.49\textwidth]{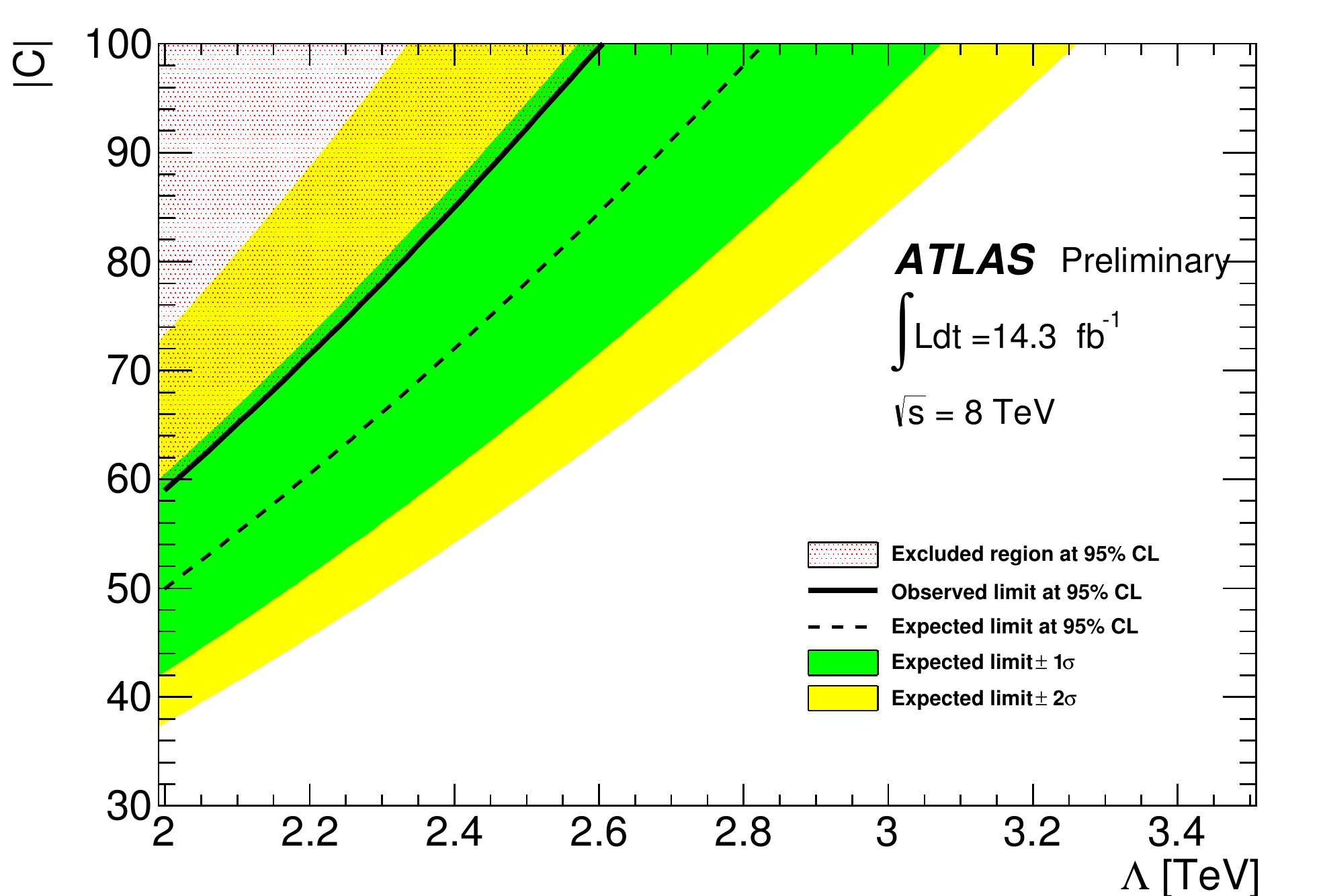}} \\
    \subfigure[]{\includegraphics[width=0.49\textwidth]{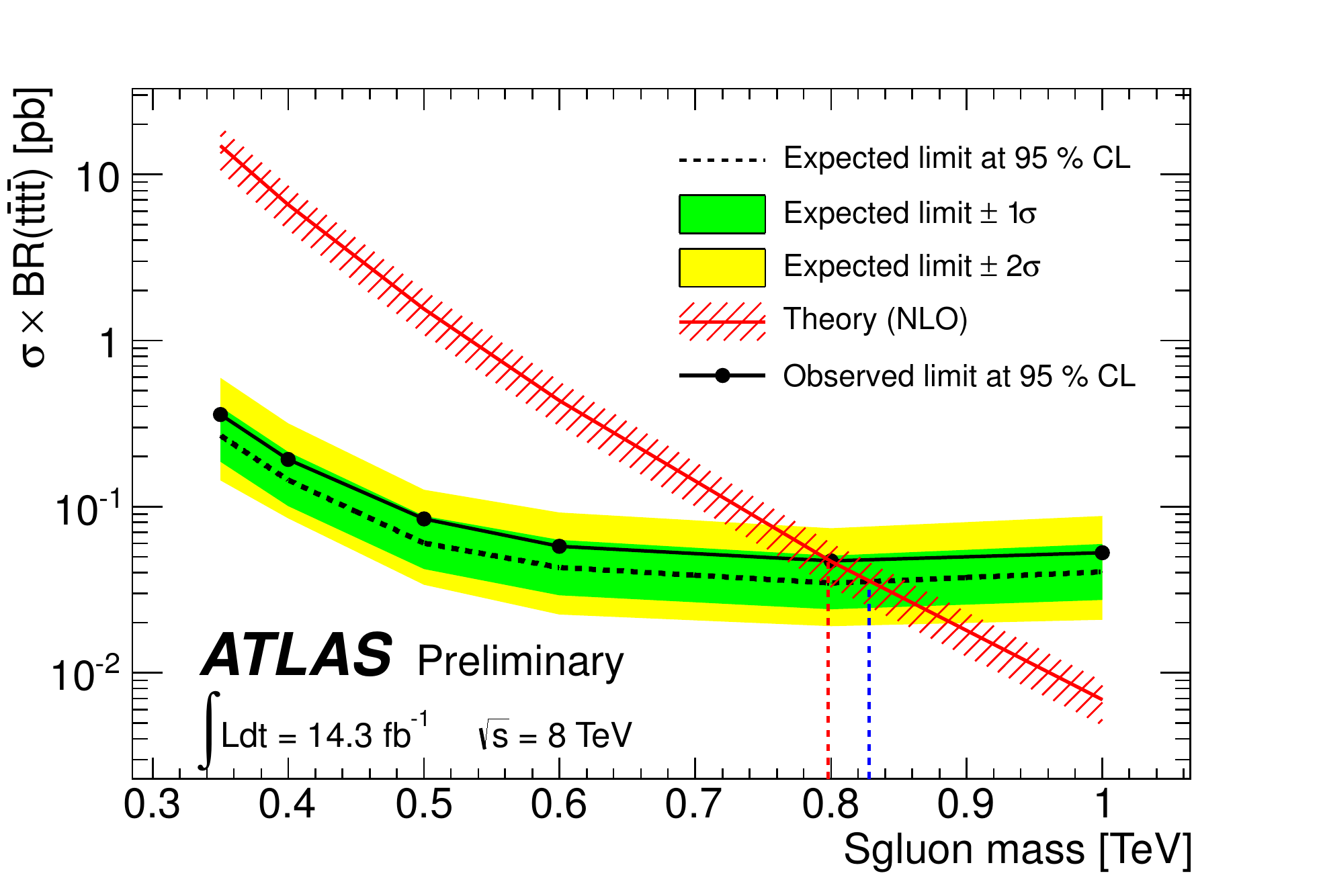}}
    \subfigure[]{\includegraphics[width=0.49\textwidth]{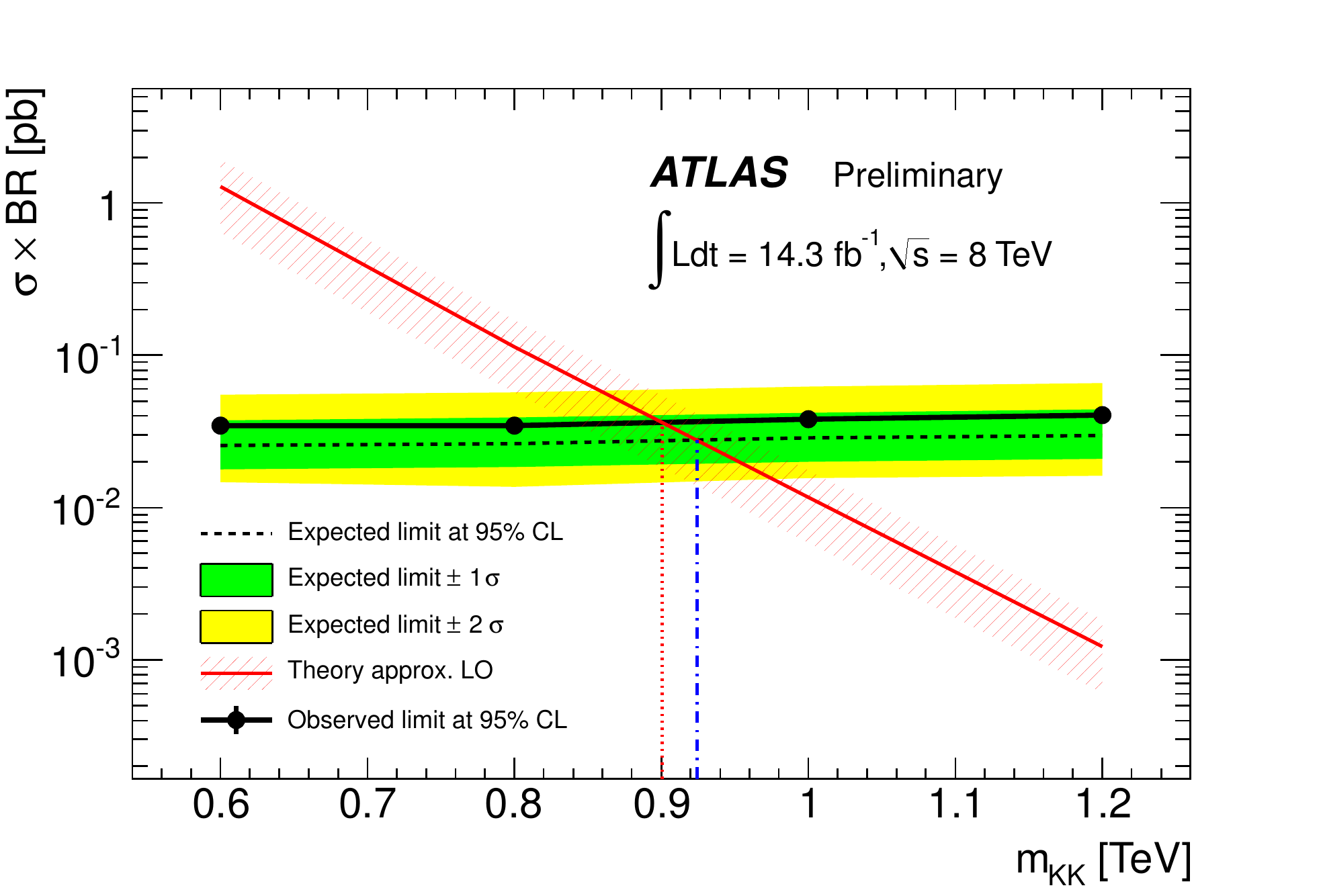}}
    \caption{Limits on the four top quarks production interpreted in the
      context of (a) the four-fermion contact interaction, (b) sgluon pair
      production, and (c) the 2UED/RPP model (in (a) the hashed region is
      excluded at 95\% C.L.).}\label{fig:limitObs:4t}
  \end{center}
\end{figure}

\section{Conclusion}

A search for new physics has been performed using events with two same-sign
leptons, at least one identified $b$ jet,  sizable \met{}, and large \HT.
The search has been done in the context of several new physics models, with
event selection criteria optimised for each scenario.  No significant excess
of events over background is observed for any of the selections, resulting
in the 95\% C.L. limits $m_{b^\prime}>0.72$~\TeV{} (assuming 100\% branching
fraction to $Wt$), $m_B>0.59$ \TeV{}, $m_T>0.54$~\TeV{} (in a natural
singlet scenario), positively-charged top pair cross section $<0.21$~pb,
sgluon mass $>0.80$~\TeV{}, and Kaluza-Klein mass $>0.90$~\TeV{}.

\end{document}